\documentclass[12pt]{revtex4}

\usepackage{graphicx,amssymb,amstext,savesym}
\usepackage{epsfig}
\usepackage{epstopdf}
\usepackage{color,graphics}

\newcommand\be{\begin{equation}}
\newcommand\eb{\end{equation}}
\newcommand\ba{\begin{eqnarray}}
\newcommand\ay{\end{eqnarray}}

\begin{document}

\title[Two-stream instability in magnetized quantum plasma]{A new two-stream instability mode in magnetized quantum plasma
}

\author{Fernando Haas}

\address{Instituto de F\'{\i}sica, Universidade Federal do Rio Grande do Sul, 91501-970, Porto Alegre, RS, Brazil}

\author{Bengt Eliasson}

\address{SUPA, Physics Department,
John Anderson Building,
Strathclyde University,
Glasgow G4 0NG,
Scotland, UK}

\begin{abstract}
A new transverse mode in a two-stream magnetized quantum plasma is studied by means of a quantum hydrodynamic model, under
non-relativistic and ideal Fermi gas assumptions. It is found that Fermi pressure effects induce a minimum cutoff wavelength for instability, unlike the classical case which is unstable for larger wavenumbers. The external magnetic field is also shown to produce a stabilizing effect. Conditions for the applicability of the model and specific parameters for experimental observations are thoroughly discussed. 
\end{abstract}


\maketitle

\section{Introduction}

The quantum two-stream instability has attracted attention since it is a benchmark problem in quantum plasmas. In Ref \cite{h1}, it has been treated by means of a multistream Schr\"odinger-Poisson model and a new quantum unstable mode was identified. The instability was shown to be due to the mode coupling involving negative-energy waves \cite{h2}. The longitudinal and transverse unstable modes have been found to be described by a generalized dispersion relation, allowing arbitrary orientation of the wave vector \cite{b}. In addition, relativistic effects reduces the growth rate of the two-stream instability, as revealed by a multistream Klein-Gordon-Poisson model \cite{h3}.

The above results are valid for the non-magnetized case. In Ref. \cite{Ren}, a non-relativistic theory of the quantum two-stream instability was considered including a homogeneous equilibrium magnetic field, and the general structure of the corresponding dispersion relation was derived. It is the purpose of the present work, to consider in detail the quantum two-stream instability in magnetized dense plasma, in a certain transverse configuration to be specified in the next Section. The peculiarity of the chosen setup is that it comprises all relevant influences in the problem, namely the Fermi pressure, the external magnetic field, the Bohm potential and the streams velocity. Therefore, a comparison of the strengths of each of these effects can be checked in detail.


This work is organized as follows. In Section II, the quantum two-stream hydrodynamic model is presented, as well as the forms of 
the two-stream equilibrium and of the small-amplitude transverse perturbations. The linear dispersion relation, the instability condition and the linear growth rate are derived and plotted for a few sets of parameters corresponding to low and high electron number densities. 
In Section III the applicability of the model in real systems is addressed.  Finally, some conclusions are drawn in Section IV.

\section{A new transverse quantum two-stream mode in magnetized plasma}
Our basic set of equations is given by the quantum hydrodynamic model for plasmas \cite{Haas},
\begin{eqnarray}
\label{e1}
\frac{\partial n_i}{\partial t} + \nabla\cdot(n_i {\bf u}_i) &=& 0 \,, \\
\label{e2}
\frac{\partial{\bf u}_i}{\partial t} + {\bf u}_{i}\cdot\nabla{\bf u}_i &=& - \frac{\nabla P_i}{m n_i} - \frac{e}{m}({\bf E} + {\bf u}_{i}\times{\bf B}) + \frac{\hbar^2}{2 m^2}\nabla\left(\frac{\nabla^{2}\sqrt{n_i}}{\sqrt{n_i}}\right) \,,\\
\label{e3}
\nabla\times{\bf E} + \frac{\partial{\bf B}}{\partial t} &=& 0 \,, \qquad
\nabla\times{\bf B} = - \mu_0 e \sum_{i=1,2} n_i {\bf u}_i + \frac{1}{c^2}\frac{\partial{\bf E}}{\partial t} \,,
\end{eqnarray}
adapted to the case of two electron streams described by number densities $n_i$ and velocity fields ${\bf u}_i$, with $i = 1,2$. Here, $m$ and $-e$ are the electron mass and charge, ${\bf E}$ and ${\bf B}$ are respectively the electric and magnetic fields, $\mu_0$ the vacuum magnetic permeability, $c$ the speed of light and $\hbar$ the scaled Planck's constant. Finally, $P_i$ is the scalar pressure of each beam, which is included since in principle the cold beam assumption can be violated, specially for very dense, degenerate streams.

Equations ({\ref{e1})-(\ref{e3}) are almost the same as those used by Stenflo \cite{Stenflo} in the study of nonlinear interactions between three ordinary mode electromagnetic waves, with two additional features. First, quantum wave-particle effects are included by means of the Bohm potential term proportional to $\hbar^2$ in Eq. (\ref{e2}). Second, quantum statistical effects arising from the Pauli exclusion principle in a dense Fermi gas are also taken into account, by means of the pressure terms. These are described by the equation of state
\begin{equation}
\label{e4}
P_i = \frac{2 n_0 E_F}{5} \left(\frac{n_i}{n_0}\right)^3  \,, \quad i = 1, 2 \,,
\end{equation}
where $n_0$ is the equilibrium number density of each stream (for simplicity we treat the symmetric case where $n_1 = n_2$ at equilibrium) and $E_F = \hbar^2 (3 \pi^2 n_0)^{2/3}/(2 m)$ is the corresponding Fermi energy. Pressure terms are in principle necessary, since for large enough densities one can have $E_F$ of the same order of the beams kinetic energy, so that the cold beam hypothesis would be unjustified. In addition, the equation of state with $P_i \sim n_{i}^{\gamma_A}$ is consistent with the case of adiabatic compression (appropriate for fast phenomena) in one spatial dimension, for which the adiabatic index is $\gamma_A = 3$. Moreover, for purely electrostatic oscillations involving just one stream, the equation of state (\ref{e4}) can be shown to reproduce the dispersion relation for high frequency waves in a fully degenerate ideal Fermi gas of electrons.
Finally we assume a not too strong magnetic field, so that an anisotropic pressure dyad is not necessary. We also assume a fixed ionic background to ensure global charge neutrality. For the fast processes to be considered, it is safe to assume immobile ions.

The equilibrium state is chosen as
\begin{equation}
\label{e5}
n_1 = n_2 = n_0 \,, \quad {\bf u}_1 = - {\bf u}_2 = u_0 \,\hat{z} \,, \quad {\bf B} = B_{0} \,\hat{z} \,, \quad {\bf E} = 0 \,,
\end{equation}
where $u_0$ is a constant speed and $B_0$ a constant magnetic field intensity. We assume transverse (to the static magnetic field) small amplitude wave perturbations proportional to $\exp[i({\bf k}\cdot{\bf r} - \omega t)]$, where ${\bf k} = k \,\hat{x}$ is the wave vector and $\omega$ the wave angular frequency. Moreover, the electric field perturbation is assumed to be along the equilibrium magnetic field direction, or $\delta{\bf E} = \delta E\, \hat{z}$, implying from Maxwell's equations a magnetic field fluctuation $\delta{\bf B} = \delta B \,\hat{y}$.

The general dispersion relation for the magnetized quantum two-stream instability has been worked out \cite{Ren}, but with no particular attention paid to the above configuration (\ref{e5}). The peculiarity of the resulting mode is that it is influenced by all relevant effects of the problem, namely Fermi pressure, magnetic field and quantum diffraction effects, as well as streaming velocities. For instance, it can be verified that for longitudinal wave propagation (${\bf k} = k \,\hat{z}$), one has: a) for electric field perturbation $\delta{\bf E}$ with $\delta{\bf E}\cdot\hat{z} \neq 0$, there is no influence of $B_0$ in the final dispersion relation; b) for $\delta{\bf E}\times\hat{z} \neq 0$ there is no role played either by the Fermi pressure or the Bohm potential. On the other hand, for transverse modes (${\bf k} = k \,\hat{x}$) and $\delta{\bf E}\times\hat{z} \neq 0$ there is no role of the streaming velocities. The equilibria mentioned in this paragraph have been discussed in the literature \cite{Ren}. On the other hand, as will be shown below, the proposed configuration (\ref{e5}) allows a comparison between the strengths of all the physical mechanisms present in the problem.


Linearizing the equations around the equilibrium (\ref{e5}) and Fourier analyzing the linearized system in space and time, the result is
\begin{equation}
\label{e6}
\omega^2 - c^2 k^2 = \omega_{p}^{2} \left(1 + \frac{k^2 u_{0}^2}{\omega^2 - 3 k^2 v_{F}^2/5 - \omega_{c}^2 - \hbar^2 k^4/(4 m^2)}\right) \,,
\end{equation}
where $\omega_p = [(2 n_0) e^2/(m\varepsilon_0)]^{1/2}$ is the plasma frequency in terms of the total electron number density $2n_0$ and the vacuum permittivity $\varepsilon_0$, $v_F = (2E_{F}/m)^{1/2}$ is the Fermi velocity and $\omega_c = e B_{0}/m$ is the gyro-frequency. The dispersion relation (\ref{e6}) agrees with Eq. (37) of \cite{Machabeli} in the special case with $v_F = 0$ and no quantum diffraction included.

The dispersion relation is a quadratic equation for $\omega^2$ that can be readily solved. In terms of the normalized variables
\begin{equation}
\label{e7}
\Omega = \frac{\omega}{\omega_p} \,, \quad K = \frac{c k}{\omega_p} \,, \quad V_F = \frac{v_F}{c} \,, \quad \Omega_c = \frac{\omega_c}{\omega_p} \,, \quad H = \frac{\hbar\omega_p}{m c^2} \,, \quad \beta = \frac{u_0}{c} \,,
\end{equation}
the result is
\begin{eqnarray}
\Omega^2 &=& \frac{1}{2}\left(1 + K^2 (1 + \frac{3}{5}V_{F}^2) + \Omega_{c}^2 + \frac{H^2 K^4}{4}\right) \nonumber \\
\label{e8}
&\pm& \frac{1}{2}\left[\left(1 + K^2 (1 - \frac{3}{5}V_{F}^2) - \Omega_{c}^2 - \frac{H^2 K^4}{4}\right)^2 + 4 \beta^2 K^2\right]^{1/2} \,.
\end{eqnarray}
As we will see below, the terms proportional to $V_F^2$, $\Omega_c^2$ and $H^2$, all have stabilizing effects on the instability.
If these terms are neglected, the instability has asymptotically the growth-rate $\Gamma=\beta$ for $K\gg 1$, where we used $\Omega=i\Gamma$.
The dominating stabilizing effect for a plasma of low density comes from the magnetic field $\sim \Omega_c$. This is the classical situation already discussed in the literature, see Ref. \cite{Machabeli} for more details, where the instability was proposed as a possible mechanism for magnetic field generation in Crab Nebula. Related works also include counter-streaming non-relativistic and relativistic magnetized plasmas \cite{Stockem06,Stockem07} with application to active galactic nuclei, etc. In the classical regime there is an instability provided $(\beta^2 - \Omega_{c}^2) K^2 > \Omega_{c}^2$. It follows that a necessary condition for instability is $\beta > \Omega_c$, i.e. in dimensional units the streaming speed must exceed the electron Alfv\'en speed $c\omega_c/\omega_p$. In strongly magnetized solid density
plasmas \cite{Tatarakis02,Wagner04}, e.g. $B_0=10^5\,\mathrm{T}$ and using solid density $n_0=10^{29}\mathrm{m}^{-3}$
we would have $\Omega_c\approx 1$ and a relativistic $\beta\approx 1$ would be required to excite the instability.
At lower magnetic fields, e.g. $B_0=10^2$-$10^3\,\mathrm{T}$ \cite{Lancia14},
we would have $\beta=10^{-2}$-$10^{-3}$.

The next important stabilizing terms for dense plasmas
comes from the Fermi pressure term proportional to $V_{F}^2$ and the Bohm pressure term proportional to $H^2$.
For solid density $n_0=10^{29}\,\mathrm{m}^{-3}$ we have $H=3.2\times10^{-5}$ and $V_F=5.5\times10^{-3}$.
The Fermi and Bohm pressure terms have equal magnitudes, $H^2 K^4/4=3 K^2 V_{F}^2/5$, when in dimensional variables the wavenumber
$k=(12/5)^{1/2}(3\pi^2)^{1/3} n_0^{1/3}\approx 4.8\, n_0^{1/3}$, which gives the
wavelength $\lambda\approx 1.3\, n_0^{-1/3}$. Hence the wavelength must be comparable to or smaller
than the mean inter-particle distance for
the Bohm potential term to dominate over the Fermi pressure term.
This is usually not possible to model within a fluid model, since the fluid model will break down at extremely small wavelengths.
Hence in our treatment we should limit the normalized wavenumber to $K\lesssim 2 V_{F}/H$
to avoid wavelengths shorter than the inter-particle distance.
Therefore, while {\it a priori} the Bohm pressure was included for the sake of completeness, {\it a posteriori} it is found that terms proportional to $H^2$ have only a marginal role
in the instability. Therefore for simplicity from now on we set $H \equiv 0$. Note that a separate kinetic, not restricted to large wavelengths treatment also indicate the presence of the Bohm contribution $\sim H^2 K^4$ \cite{Eliasson10}. However, in the practical applications to be described below the Fermi pressure terminates the instability well before the Bohm term starts to play a major role.

Taking the negative sign in Eq. (\ref{e8}) can give rise to instability ($\Omega^2 < 0$), provided
\begin{equation}
\label{e9}
\frac{\beta^2 K^2}{1+K^2} > \Omega_{c}^2 + \frac{3}{5}K^2 V_{F}^2 \,.
\end{equation}
With $\Omega = i \Gamma$, the resulting growth rate would follows from
\begin{eqnarray}
\Gamma^2 &=& \frac{1}{2}\left[\left(1 + K^2 (1 - \frac{3}{5}V_{F}^2) - \Omega_{c}^2\right)^2 + 4 \beta^2 K^2\right]^{1/2}
\nonumber \\
\label{e10}
&-& \frac{1}{2}\left(1 + K^2 (1 + \frac{3}{5}V_{F}^2) + \Omega_{c}^2\right) \,.
\end{eqnarray}

A necessary requirement to fulfill the inequality (\ref{e9}) is $\beta > \Omega_c$, as can be verified.
It is apparent that both magnetic field and Fermi pressure effects are stabilizing, due to the terms proportional resp. to $\Omega_{c}^2$ and $V_{F}^2$ terms in the instability condition. From the same contributions, it is found that both for $K \rightarrow 0$ and $K \rightarrow \infty$ one has stable modes. Without Fermi pressure, $K \rightarrow \infty$ is unstable assuming $\beta > \Omega_c$.

The instability condition (\ref{e9}) can be treated in exact analytical form since it correspond to a second-degree equation for $K^2$. However, the resulting expressions are somewhat cumbersome, so that we prefer to single out the two limiting relevant subcases below.

\subsection{Classical case, $\Omega_c \gg K V_F$}
This is the situation already discussed in the literature, see Ref. \cite{Machabeli} for more details, and also the related works
in Refs. \cite{Stockem06,Stockem07}. In the classical regime there is instability provided $(\beta^2 - \Omega_{c}^2) K^2 > \Omega_{c}^2$. It follows that a necessary condition for instability is $\beta>\Omega_c$, i.e. in dimensional units the streaming speed must exceed the electron Alfv\'en speed $c\omega_c/\omega_p$.

\subsection{Quantum dominated case, $K V_F \gg \Omega_c$}
In this regime, the instability condition (\ref{e9}) simplifies to
\begin{equation}
\label{e11}
K^2 < \frac{5}{3}\frac{\beta^2}{V_{F}^2} - 1 \,,
\end{equation}
%
%
%
requiring
\begin{equation}
\label{e12}
\beta^2 > \frac{3}{5} V_{F}^2
\end{equation}
and predicting a cutoff wavenumber
\begin{equation}
\label{e13}
K_c = \left(\frac{5}{3}\frac{\beta^2}{V_{F}^2} - 1\right)^{1/2}
\end{equation}
above which there is no instability anymore. Re-introducing dimensional variables one has a cutoff wavelength
\begin{equation}
\label{e14}
\lambda_c = \frac{2\pi c}{\omega_p} \left(\frac{5}{3}\frac{u_{0}^2}{v_{F}^2} - 1\right)^{-1/2} \,,
\end{equation}
so that $\lambda < \lambda_c$ are certainly stable.

\begin{figure}
\centering
\includegraphics[width=12cm]{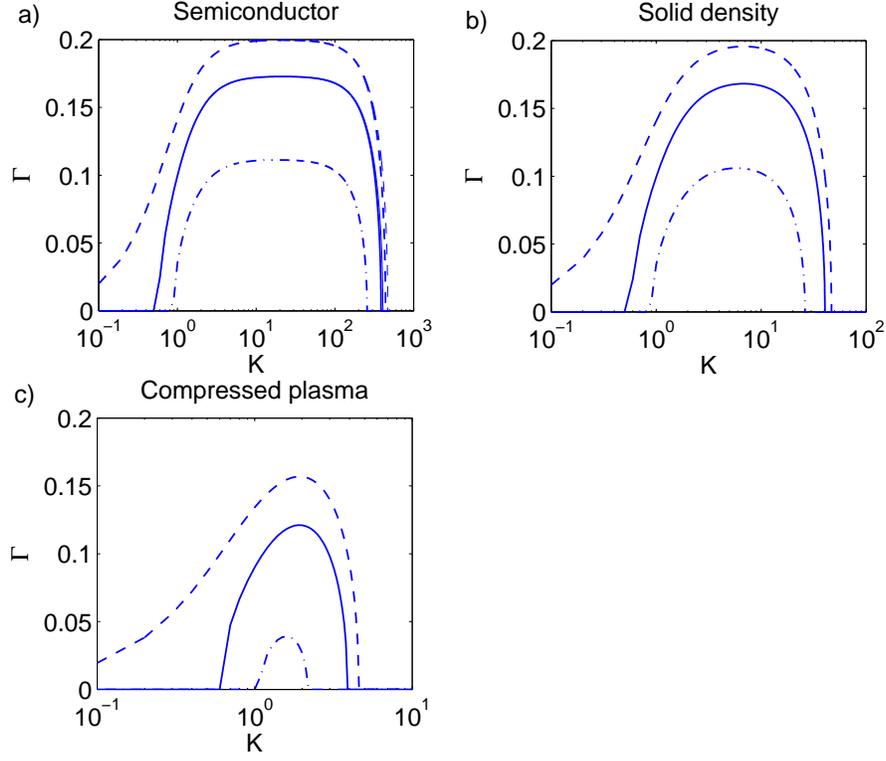}
\vskip 5cm
\caption{Growth-rates for different sets of parameters corresponding to a) a semiconductor plasma with number density $n_0=10^{26}\,\mathrm{m}^{-3}$ corresponding to $V_F=5.5\times10^{-4}$ and $H=1.03\times10^{-6}$,
b) a solid density plasma with $n_0=10^{29}\,\mathrm{m}^{-3}$ corresponding to $V_F=5.5\times10^{-3}$ and $H=3.24\times10^{-5}$, and c) a compressed
plasma with $n_0=10^{32}\,\mathrm{m}^{-3}$ corresponding to $V_F=5.5\times10^{-2}$ and $H=1.03\times10^{-3}$.
The velocities and magnetic field are chosen such that $\beta=0.2$ and $\Omega_c=0.1$ (solid lines), $\beta=0.2$ and $\Omega_c=0$ (dashed lines), and $\beta=0.15$ and $\Omega_c=0.1$ (dash-dotted lines). In all cases there is only a marginal influence of the Bohm pressure terms proportional to $H^2$, while the Fermi pressure term proportional to $V_F^2$ terminates the instability at large $K$. As an illustration,
thinner lines show the case $H=0$ in panel a) showing a barely visible effect at large wavenumbers.
}
\end{figure}

Figure 1 shows the growth-rate obtained from Eq. (8), including $H$ for the sake of completeness, as a function of wavenumber for different sets of parameters corresponding to a low-density
semiconductor plasma with a density of $n_0=10^{26}\,\mathrm{m}^{-3}$ (corresponding to $V_F=5.5\times10^{-4}$ and $H=1.03\times10^{-6}$), a solid density plasma $n_0=10^{29}\,\mathrm{m}^{-3}$ (corresponding to $V_F=5.5\times10^{-3}$ and $H=3.24\times10^{-5}$), and a compressed
plasma with $n_0=10^{32}\,\mathrm{m}^{-3}$ (corresponding to $V_F=5.5\times10^{-2}$ and $H=1.03\times10^{-3}$).
The magnetic field has a stabilizing effect and gives rise to a smallest unstable wavenumber. As mentioned above the velocity
$\beta$ must be larger than $\Omega_c$ for instability.
The instability in general is larger for larger $\beta$ and is terminated at large wavenumbers due to the
Fermi pressure proportional to $V_F^2$. The Bohm pressure
is important only at very large wavenumbers $K\sim V_F/H$ and plays only a minor role for the used parameters since the instability takes place only for $K < V_F/H$.

In brief, the results of this Section indicate that: (a) the effects of the Bohm pressure are relatively unimportant for the specific transverse mode under consideration; (b) magnetic and Fermi pressure effects are stabilizing and can suppress the instability, whenever $\Omega_c > \beta$ or $\lambda < \lambda_c$; (c) in addition, precise statements about the instability have been provided, in particular the existence condition (\ref{e9}) and the growth rate from Eq. (\ref{e10}). In the next Section, there is a discussion on the practical aspects of the predictions, regarding possible experimental verification.

\section{Observability issues}
The obvious questions about the above results are: (a) is the theory applicable to the real world? (b) which parameters should be applied in eventual laboratory (or astrophysical) environments? We have to address these questions in this Section. As will become  apparent, a series of conflicting requirements shows up, but a suitable numerical range of physical parameters are proposed.

First of all, there is no collisional mechanism in the model equations (\ref{e1})-(\ref{e2}). However, the ideal Fermi gas assumption tends to be more justifiable for sufficiently dense systems. Indeed, in this case the exclusion principle triggers the Pauli blocking mechanism
forbidding transitions between electrons of the same quantum state \cite{Ashcroft}, effectively implying weaker local interactions (collisions) between the charge carriers. Moreover, the most relevant non-ideal aspect, namely electron-electron collisions, has a collisional frequency $\nu_{ee}$ which can be estimated \cite{Ashcroft} by
\begin{equation}
\label{e15}
\hbar\nu_{ee} \approx E_F \, \left(\frac{T}{T_F}\right)^2 \sim n_{0}^{-2/3} \,,
\end{equation}
where $T$ is the thermodynamic temperature and $T_F = E_{F}/\kappa_B$ is the Fermi temperature, in terms of the Boltzmann constant
$\kappa_B$. The right hand side is smaller for larger density and smaller thermodynamic temperature. In particular, a fully degenerate system ($T \rightarrow 0$) can be safely assumed as ideal. For practical issues, the damping rate resulting from Eq. (\ref{e15}) should be compared to the growth rate from Eq. (\ref{e10}). Finally, even if denser systems tend to be more ideal, a non-relativistic model requires $v_{F}/c \ll 1$, which is safe for number densities lower than $n_0 \approx 10^{36} m^{-3}$ (a typical number density in the interior of white dwarf stars). Otherwise, relativistic corrections will come into play.

In addition, for the cutoff wavelength predicted in Eq. (\ref{e14}) one expects that
\begin{equation}
\label{e16}
\lambda_c \gg \lambda_0 \approx n_{0}^{-1/3} \,,
\end{equation}
where the right-hand side denotes $\lambda_0$ as a measure of the average inter-particle distance. Otherwise, the application of a
fluid modeling may not be valid. The condition (\ref{e16}) can be worked out to give
\begin{equation}
\label{e17}
n_{0}^{1/3} \gg \frac{\omega_p}{2\pi c} \left(\frac{5}{3}\frac{u_{0}^2}{v_{F}^2} - 1\right)^{1/2} \,.
\end{equation}
%
%
%
Since $\omega_p \sim n_{0}^{1/2}$, Eq. (\ref{e17}) is fulfilled for not too high number densities. 
This is also required in view of the inequality (\ref{e12}), since a large Fermi speed is able to arrest the instability.

A further requirement to apply a quantum fluid theory is that $k \lambda_F \ll 1$, where $\lambda_F = v_{F}/(\sqrt{3}\,\omega_p)$ is the Thomas-Fermi screening length \cite{Haas}, playing the analog role, in degenerate plasma, of the Debye length in classical plasma. Molecular simulation of Yukawa quantum fluids \cite{Schmidt} provide support to this long wavelength assumption. In the present problem, an equivalent form of it is $\lambda_c \gg \lambda_F$, which can be worked out as
\begin{equation}
\label{e19}
v_{F}^2 + 12 \pi^2 c^2 \gg \frac{5}{3}u_{0}^2 \,,
\end{equation}
which is automatically fulfilled in view of the non-relativistic assumption.

As an illustration and for definiteness, we can chose a setup with $u_0 = c/10$ and a density $n_0 = 10^{31} m^{-3}$, which in principle is accessible \cite{Nuckolls72,Azechi91,Kodama01} in present day laser-plasma compression experiments. In this case we have $v_{F}/c = 0.03, \lambda_c = 1.52\, nm$ (in the soft X-ray range) and $\lambda_c/\lambda_0 = 32.78$, which are in accordance with the previous analysis. Also note that for this arrangement one has $\omega_p = 2.52 \times 10^{17} s^{-1}$ and $T_F = 1.97 \times 10^6 K$, so that $T \ll 10^6 K$ would be needed to justify the full degeneracy assumption.

Additionally, it has been shown that the instability condition can be satisfied only if $\beta > \Omega_c$, which in this case yields $B_0 < 1.44 \times 10^5 \,T$. The external magnetic field can be regarded as a control parameter to further validate the theory, e.g. switching it on until the instability stops.

The ideality condition also needs to be verified. From Eq. (\ref{e15}) and the elected parameters, one has a damping rate $\nu_{ee} = 6.65 \times 10^4 \,T^2$ and a maximum growth rate (as numerically found for $B_0 = 0$) given by $\gamma_{max} = 0.08\, \omega_p$, a fast enough instability to overcome collisional damping. For instance, for $T = 10^4 K \ll T_F$ one has $\nu_{ee}/\gamma_{max} = 3.30 \times 10^{-4}$. A critical experimental issue, among others, would be to maintain a small thermodynamic  temperature of each electron stream.

To conclude the Section, in the suggested setup we have $\lambda_{c}/\lambda_F = 86.02$, which is consistent with a 
fluid description of the Fermi gas.

\section{Conclusion}
A transverse unstable mode in the non-relativistic quantum two-stream instability in magnetized dense plasmas was analyzed in detail. The Fermi pressure was shown to be the dominant quantum effect in such degenerate plasmas, providing a mechanism for the arrest of the instability for sufficiently small wavelengths, besides the classical stabilizing role of the external magnetic field. The physical parameters for possible experiments have been worked out. Possible extensions could be the investigation of oblique modes, or the inclusion of relativistic effects.


{\bf Acknowledgments}

 F. H. acknowledges the Brazilian research funding agency CNPq (Conselho Na\-cio\-nal de De\-sen\-volvimento Cient\'{\i}fico e Tecnol\'ogico) for financial support. B. E. acknowledges support from the Engineering and Physical
Sciences Research Council (EPSRC), U.K., Grant no. EP/M009386/1.

\end{document}